\documentclass[aps,nofootinbib,preprintnumbers]{revtex4}
\usepackage{lipsum}
\usepackage{CJK}
\usepackage{amsfonts}
\usepackage{amsmath}
\usepackage{amssymb}
\usepackage[english]{babel}
\usepackage{graphicx}
\usepackage{epsfig}
\usepackage{bm}
\usepackage{verbatim}
\usepackage[utf8]{inputenc}
\usepackage{booktabs}
\usepackage{multirow}
\usepackage{subfig}
\usepackage{slashed}
\usepackage{xcolor}
\usepackage[colorlinks=true,urlcolor=red,citecolor=red]{hyperref}
\usepackage[font=small]{caption}
\usepackage{float}
\usepackage{blindtext}
\usepackage{placeins}
\usepackage[utf8]{inputenc}
\usepackage{bbm}
\usepackage[colorinlistoftodos]{todonotes}

\newenvironment{Eqnarray}{\arraycolsep 0.14em\begin{eqnarray}}{\end{eqnarray}}
\newcommand{\ba}{\begin{Eqnarray}}
\newcommand{\ea}{\end{Eqnarray}}
\newcommand{\be}{\begin{equation}}
\newcommand{\ee}{\end{equation}}
\newcommand{\bal}{\begin{aligned}}
\newcommand{\eal}{\end{aligned}}
\newcommand{\bea}{\begin{eqnarray}}
\newcommand{\eea}{\end{eqnarray}}
\newcommand{\ben}{\begin{enumerate}}
\newcommand{\een}{\end{enumerate}}
\newcommand{\bit}{\begin{itemize}}
\newcommand{\eit}{\end{itemize}}
\newcommand{\bde}{\begin{widetext}}
\newcommand{\ede}{\end{widetext}}

\def\lsim{\mathrel{\rlap{\lower4pt\hbox{\hskip1pt$\sim$}}
    \raise1pt\hbox{$<$}}}
\def\gsim{\mathrel{\rlap{\lower4pt\hbox{\hskip1pt$\sim$}}
    \raise1pt\hbox{$>$}}}

\def\3211{$\mathrm{SU(3) \otimes SU(2)_L \otimes U(1)_R \otimes U(1)_{B-L}}$ }
\def\321{$\mathrm{SU(3) \otimes SU(2) \otimes U(1)}$ }
\def\422{$\mathrm{SU(4) \otimes SU(2) \otimes SU(2)_R}$ }

\newcommand{\mathsym}[1]{{}}

\definecolor{bostonuniversityred}{rgb}{0.8, 0.0, 0.0}

\input{tcilatex}
\begin{document}

\title{An $A_5$ inverse seesaw model with perturbed golden ratio mixing}
\author{A. E. C\'arcamo Hern\'andez$^{a,b,c}$}
\email{antonio.carcamo@usm.cl}
\affiliation{\hspace{2cm}\\
$^{{a}}$Universidad T\'ecnica Federico Santa Mar\'{\i}a, Casilla 110-V,
Valpara\'{\i}so, Chile\\
$^{{b}}$Centro Cient\'{\i}fico-Tecnol\'ogico de Valpara\'{\i}so, Casilla
110-V, Valpara\'{\i}so, Chile\\
$^{{c}}$Millennium Institute for Subatomic Physics at the High-Energy
Frontier, SAPHIR, Calle Fern\'andez Concha No 700, Santiago, Chile}
\author{Ivo de Medeiros Varzielas}
\email{ivo.de@udo.edu}
\affiliation{CFTP, Departamento de F\'{\i}sica, Instituto Superior T\'{e}cnico,
Universidade de Lisboa, Avenida Rovisco Pais 1, 1049 Lisboa, Portugal}
\date{\today }
\date{\today }

\begin{abstract}
We propose a theory based on the $A_5$ discrete group which successfully
accounts for lepton masses and mixings. In the proposed model, the neutrino
masses arise from an inverse seesaw mechanism. The leptonic mixing consists
in a perturbed golden ratio mixing pattern with a strong correlation arising
between the leptonic mixing observables, which can be expressed as $%
\theta_{13}$ predicting the other mixing angles and the Dirac CP phase.
\end{abstract}

\maketitle


\section{Introduction}

\label{intro}

The problem of fermion masses and mixing is one of the theoretical issues
that afflicts the Standard Model. The typical solution to the problem is the
introduction of extra symmetries - often referred to as family symmetries -
which justify the replication of the fermions in 3 generations. Through the
mechanism of breaking the family symmetry it is possible to explain the
different and hierarchical masses of the generations and the mixing between
fermions.

The family symmetry introduced can be Abelian or non-Abelian, discrete or
continuous. Non-Abelian discrete symmetries gained favour as they are better
suited to explain leptonic mixing patterns. Within the more popular
non-Abelian discrete family symmetries in the literature are $S_3$ \cite%
{Gerard:1982mm,Kubo:2003iw,Kubo:2003pd,Kobayashi:2003fh,Chen:2004rr,Mondragon:2007af}%
, $A_{4}$ \cite{Ma:2001dn,Altarelli:2005yp,
deMedeirosVarzielas:2005qg,CarcamoHernandez:2019kjy}, $S_{4}$ \cite%
{Patel:2010hr,Morisi:2011pm,Mohapatra:2012tb,BhupalDev:2012nm,Varzielas:2012pa,VanVien:2015xha,CarcamoHernandez:2019eme,CarcamoHernandez:2019iwh}%
, $\Delta(27)$ \cite{Branco:1983tn,deMedeirosVarzielas:2006fc,Ma:2006ip,
deMedeirosVarzielas:2015amz,CarcamoHernandez:2020udg,CarcamoHernandez:2021osw}
and $A_5$ \cite%
{Everett:2008et,Feruglio:2011qq,Cooper:2012bd,Varzielas:2013hga,Gehrlein:2014wda,DiIura:2015kfa,Ballett:2015wia,Turner:2015uta,Li:2015jxa}%
. The largest of these, $A_5$, can be broken into subgroups which lead to
what is referred to as golden ratio mixing patterns.

The original golden ratio mixing pattern with vanishing $\theta_{13}$ is
clearly excluded, with the more recent $A_5$ models \cite%
{Feruglio:2011qq,Cooper:2012bd,Varzielas:2013hga,Gehrlein:2014wda,DiIura:2015kfa,Ballett:2015wia,Turner:2015uta,Li:2015jxa}
obtaining deviations via corrections in the golden mixing pattern arising
from the vacuum alignments of the $A_5$ triplets and $A_5$ quintuplet \cite%
{Feruglio:2011qq} as well as corrections in the neutrino \cite%
{Cooper:2012bd,DiIura:2015kfa,Ballett:2015wia,Turner:2015uta,Li:2015jxa} or
charged lepton \cite{Cooper:2012bd,Gehrlein:2014wda} sectors. In this paper
we explore controlled deviations of the golden ratio mixing patterns, in
order to have an excellent fit to leptonic mixing data. We consider an
extended fermion sector leading to an inverse seesaw mechanism, such that,
on the family symmetry basis, the neutrino mixing is the Golden Ratio, but
on the same basis the charged lepton mass is not diagonal. The leptonic
mixing therefore is a perturbed Golden Ratio pattern.

In Section \ref{sec:model} we describe the model in terms of symmetries and
the respective assignments of the fields. Section \ref{sec:massmix} details
the breaking of the family symmetry and the associated mass terms that
arise, leading to the leptonic masses and mixing. We conclude in Section \ref%
{sec:conc}.

\section{The model \label{sec:model}.}

The model under consideration is an $A_{5}$ flavor supersymmetric theory for
leptons where the inverse seesaw mechanism is implemented to generate the
tiny masses of the light active neutrinos and lepton mixings features a
perturbed golden ratio pattern. The $A_{5}$ family symmetry of the model is
supplemented by the $Z_{9}\times Z_{N}$ discrete group. The successfull and
viable implementation of the inverse seesaw mechanism requires the inclusion
of six right handed Majorana neutrinos $\nu _{iR}$ and $N_{iR}$ ($i=1,2,3$),
which are grouped into an $A_{5}$ leptonic triplet. Furthermore, in the
supersymmetric model under consideration, the scalar sector of the MSSM is
extended by the inclusion of several gauge singlet scalar fields. Such gauge
singlet scalar fields are needed to generate the required Yukawa terms in
the neutrino and charged lepton sectors crucial for a successfull
implementation of the inverse seesaw mechanism and to generate the SM
charged lepton masses as well as perturbed golden ratio pattern of lepton
mixing consistent with the experimental data. We use a SUSY framework
instead of a non SUSY one, since the former allows to naturally obtain the
VEV configuration for the $A_{5}$ scalar triplets and the $A_{5}$ scalar
quintuplet (to be shown below) that yields a perturbed golden mixing pattern
for the leptonic mixings. The leptonic and scalar assignments under the $%
A_{5}\times Z_{9}\times Z_{N}$ discrete group are shown in Table \ref%
{leptons}. 
\begin{table}[tbp]
\begin{tabular}{|c|c|c|c|c|c|c|c|c|c|c|c|c|c|c|c|c|}
\hline
& $l_{L}$ & $l_{1R}$ & $l_{2R}$ & $l_{3R}$ & $\nu _{R}$ & $N_{R}$ & $H_{u}$
& $H_{d}$ & $\varphi _{e}$ & $\varphi _{\mu }$ & $\varphi _{\tau }$ & $%
\theta $ & $\rho $ & $\varphi _{N}$ & $\xi _{1}$ & $\xi _{2}$ \\ \hline
$A_{5}$ & $\mathbf{\mathbf{3}}$ & $\mathbf{\mathbf{1}}$ & $\mathbf{\mathbf{1}%
}$ & $\mathbf{\mathbf{1}}$ & $\mathbf{\mathbf{3}}$ & $\mathbf{\mathbf{3}}$ & 
$\mathbf{\mathbf{1}}$ & $\mathbf{\mathbf{1}}$ & $\mathbf{\mathbf{3}}$ & $%
\mathbf{\mathbf{3}}$ & $\mathbf{\mathbf{3}}$ & $\mathbf{\mathbf{1}}$ & $%
\mathbf{\mathbf{1}}$ & $\mathbf{\mathbf{5}}$ & $\mathbf{\mathbf{1}}$ & $%
\mathbf{\mathbf{1}}$ \\ \hline
$Z_{9}$ & $-1$ & $-4$ & $-3$ & $-2$ & $-1$ & $2$ & $0$ & $1$ & $0$ & $0$ & $%
0 $ & $0$ & $1$ & $3$ & $3$ & $-3$ \\ \hline
$Z_{N}$ & $0$ & $0$ & $-x$ & $x$ & $0$ & $0$ & $0$ & $0$ & $0$ & $x$ & $-x$
& $-x$ & $0$ & $0$ & $0$ & $0$ \\ \hline
\end{tabular}%
\caption{Lepton and scalar assigments under the $A_{5}\times Z_{9}\times
Z_{N}$ discrete group.}
\label{leptons}
\end{table}
With the particle content and symmetries specified in Table \ref{leptons},
the following charged lepton and neutrino Yukawa interactions arise: 
\begin{eqnarray}
-\mathcal{L}_{Y}^{\left( l\right) } &=&y_{1}^{\left( l\right) }\left( 
\overline{l}_{L}H_{d}\varphi _{e}\right) _{\mathbf{\mathbf{1}}}l_{1R}\frac{%
\rho ^{2}}{\Lambda ^{3}}+y_{2}^{\left( l\right) }\left( \overline{l}%
_{L}H_{d}\varphi _{\mu }\right) _{\mathbf{\mathbf{1}}}l_{2R}\frac{\rho }{%
\Lambda ^{2}}+y_{3}^{\left( l\right) }\left( \overline{l}_{L}H_{d}\varphi
_{\tau }\right) _{\mathbf{\mathbf{1}}}l_{3R}\frac{1}{\Lambda }  \notag \\
&&+z_{1}^{\left( l\right) }\left( \overline{l}_{L}H_{d}\varphi _{\mu
}\right) _{\mathbf{\mathbf{1}}}l_{1R}\frac{\theta \rho ^{2}}{\Lambda ^{4}}%
+z_{3}^{\left( l\right) }\left( \overline{l}_{L}H_{d}\varphi _{e}\right) _{%
\mathbf{\mathbf{1}}}l_{3R}\frac{\theta }{\Lambda ^{2}}+H.c.  \label{Lyl}
\end{eqnarray}%
\begin{equation}
-\mathcal{L}_{Y}^{\left( \nu \right) }=y_{\nu }\left( \overline{l}%
_{L}H_{u}\nu _{R}\right) _{\mathbf{\mathbf{1}}}+y_{\nu N}\left( \overline{%
\nu }_{R}N_{R}^{C}\right) _{\mathbf{\mathbf{1}}}\rho +y_{N}\left( \overline{N%
}_{R}N_{R}^{C}\right) _{\mathbf{5}}\varphi _{N}\frac{H_{u}H_{d}}{\Lambda ^{2}%
}+H.c.  \label{Lynu}
\end{equation}%
In order to get the light active neutrino mass matrix consistent with the
golden ratio mixing pattern of lepton mixing and to successfully generate
the reactor mixing angles from the charged lepton sector, we consider the
following VEV configuration for the $A_{5}$ scalar triplets and the $A_{5}$
scalar quintuplet: 
\begin{eqnarray}
\left\langle \varphi _{e}\right\rangle &=&v_{\varphi _{e}}\left(
1,0,0\right) ,\hspace{1cm}\left\langle \varphi _{\mu }\right\rangle
=v_{\varphi _{\mu }}\left( 0,1,0\right) ,\hspace{1cm}\left\langle \varphi
_{\tau }\right\rangle =v_{\varphi _{\tau }}\left( 0,0,1\right) ,  \notag \\
\left\langle \varphi _{N}\right\rangle &=&v_{\varphi _{N}}\left( -\sqrt{%
\frac{2}{3}}\left( p+q\right) ,-p,q,q,p\right) ,  \label{A5VEVs}
\end{eqnarray}%
As shown in detail in \cite{Feruglio:2011qq}, the above given VEV pattern
for the $A_{5}$ scalar triplets and the $A_{5}$ scalar quintuplet can be
obtained from the following superpotential: 
\begin{equation}
W=M_{0}\xi _{1}\xi _{2}+\kappa _{1}\xi _{1}\varphi _{N}^{2}+\kappa
_{2}\left( \varphi _{N}^{3}\right) _{1}+\kappa _{3}\left( \varphi
_{N}^{3}\right) _{2}+\frac{\kappa _{4}}{3}\xi _{1}^{3}+\frac{\kappa _{5}}{3}%
\xi _{2}^{3}+M_{1}\varphi _{e}^{2}+M_{2}\varphi _{\mu }\varphi _{\tau
}+\kappa \varphi _{e}\varphi _{\mu }\varphi _{\tau }+\cdots
\end{equation}%
where $\xi _{1}$ and $\xi _{2}$ are driving fields. Notice that the $\varphi
_{e}^{3}$ and $\varphi _{e}\varphi _{\mu }\theta $ terms are not present in
the superpotencial since they yield vanishing results as follows from the $%
A_{5}$ multiplications rules. Finally, to close this section, we briefly
compare our model with the one proposed in \cite{Feruglio:2011qq}. In the
model of \cite{Feruglio:2011qq}, the light active neutrino masses are
generated from a type I seesaw mechanism and the reactor mixing angle arises
from a perturbation of the VEV configuration of the $A_{5}$ triplets and $%
A_{5}$ quintuplet. On the other hand, in our proposed model, the masses of
the light active neutrinos are produced via an inverse seesaw mechanism and
the required perturbation of the golden mixing pattern that generates a non
vanishing reactor mixing angle arises from the charged lepton sector.

\section{Lepton masses and mixings \label{sec:massmix}}

After the SM gauge symmetry and the $A_{5}\times Z_{9}\times Z_{N}$ discrete
group are spontaneously broken, the following mass matrix for SM charged
leptons is obtained: 
\begin{equation}
M_{l}=\left( 
\begin{array}{ccc}
y_{1}^{\left( l\right) }\frac{v_{\varphi _{e}}v_{\rho }^{2}}{\Lambda ^{3}}%
\frac{v_{d}}{\sqrt{2}} & 0 & z_{3}^{\left( l\right) }\frac{v_{\varphi _{e}}}{%
\Lambda }\frac{v_{d}}{\sqrt{2}} \\ 
z_{1}^{\left( l\right) }\frac{v_{\varphi _{\mu }}v_{\rho }^{2}}{\Lambda ^{3}}%
\frac{v_{d}}{\sqrt{2}} & y_{2}^{\left( l\right) }\frac{v_{\varphi _{\mu
}}v_{\rho }}{\Lambda ^{2}}\frac{v_{d}}{\sqrt{2}} & 0 \\ 
0 & 0 & y_{3}^{\left( l\right) }\frac{v_{\varphi _{\tau }}}{\Lambda }\frac{%
v_{d}}{\sqrt{2}}%
\end{array}%
\right) ,
\end{equation}%
where we assume that the entries of $M_{l}$ fulfil the following hierarchy: 
\begin{equation}
\left\vert \left( M_{l}\right) _{11}\right\vert \sim \left\vert \left(
M_{l}\right) _{21}\right\vert <<\left\vert \left( M_{l}\right)
_{22}\right\vert \sim \left\vert \left( M_{l}\right) _{13}\right\vert
<\left\vert \left( M_{l}\right) _{33}\right\vert .
\end{equation}%
This hierarchy is justified by the insertions of the fields (such as $\rho$%
). The matrix $M_{l}M_{l}^{\dagger }$ is diagonalized by a rotation matrix
according to: 
\begin{equation}
R_{l}^{\dagger }M_{l}M_{l}^{\dagger }R_{l}=\left( 
\begin{array}{ccc}
m_{e} & 0 & 0 \\ 
0 & m_{\mu } & 0 \\ 
0 & 0 & m_{\tau }%
\end{array}%
\right) ,\hspace{1cm}R_{l}=\left( 
\begin{array}{ccc}
\cos \theta & 0 & \sin \theta e^{i\theta } \\ 
0 & 1 & 0 \\ 
-\sin \theta e^{-i\theta } & 0 & \cos \theta%
\end{array}%
\right) ,\hspace{1cm}\tan \theta \simeq \frac{\left\vert \left( M_{l}\right)
_{13}\right\vert }{\left\vert \left( M_{l}\right) _{33}\right\vert },
\end{equation}

Furthermore, the neutrino Yukawa terms yield the following neutrino mass
terms: 
\begin{equation}
-\mathcal{L}_{mass}^{\left( \nu \right) }=\frac{1}{2}\left( 
\begin{array}{ccc}
\overline{\nu _{L}^{C}} & \overline{\nu _{R}} & \overline{N_{R}}%
\end{array}%
\right) M_{\nu }\left( 
\begin{array}{c}
\nu _{L} \\ 
\nu _{R}^{C} \\ 
N_{R}^{C}%
\end{array}%
\right) +\sum_{n=1}^{2}\sum_{m=1}^{2}\left( m_{\Omega }\right) _{nm}%
\overline{\Omega }_{nR}\Omega _{mR}^{C}+H.c.,
\end{equation}
where the neutrino mass matrix is given by: 
\begin{equation}
M_{\nu }=\left( 
\begin{array}{ccc}
0_{3\times 3} & m_{\nu D} & 0_{3\times 3} \\ 
m_{\nu D}^{T} & 0_{3\times 3} & M \\ 
0_{3\times 3} & M^{T} & \mu%
\end{array}%
\right) ,  \label{Mnu}
\end{equation}
and the submatrices $m_{\nu D}$, $M$ and $\mu $ have the following structure:

\begin{equation}
m_{\nu D}=\frac{y_{\nu }v_{u}}{\sqrt{2}}\left( 
\begin{array}{ccc}
1 & 0 & 0 \\ 
0 & 0 & 1 \\ 
0 & 1 & 0%
\end{array}%
\right) ,\hspace{1cm}M=y_{\nu N}v_{\rho }\left( 
\begin{array}{ccc}
1 & 0 & 0 \\ 
0 & 0 & 1 \\ 
0 & 1 & 0%
\end{array}%
\right) ,\hspace{1cm}\mu =\frac{v_{u}v_{d}v_{\varphi _{N}}}{\sqrt{6}\Lambda
^{2}}\left( 
\begin{array}{ccc}
\frac{2}{3}\left( p+q\right) & \frac{p}{\sqrt{2}} & \frac{p}{\sqrt{2}} \\ 
\frac{p}{\sqrt{2}} & q & -\frac{1}{3}\left( p+q\right) \\ 
\frac{p}{\sqrt{2}} & -\frac{1}{3}\left( p+q\right) & q%
\end{array}%
\right) .  \label{Mnublocks0}
\end{equation}
The active light neutrino masses are generated from an inverse seesaw
mechanism, and the physical neutrino mass matrices are given by: 
\begin{eqnarray}
\widetilde{M}_{\nu } &=&m_{\nu D}\left( M^{T}\right) ^{-1}\mu M^{-1}m_{\nu
D}^{T},\hspace{0.7cm}  \label{M1nu} \\
M_{\nu }^{\left( -\right) } &=&-\frac{1}{2}\left( M+M^{T}\right) +\frac{1}{2}%
\mu ,\hspace{0.7cm} \\
M_{\nu }^{\left( +\right) } &=&\frac{1}{2}\left( M+M^{T}\right) +\frac{1}{2}%
\mu .  \label{neutrino-mass}
\end{eqnarray}
here $\widetilde{M}_{\nu }$ is the mass matrix for the active light
neutrinos ($\nu _{a}$), whereas $M_{\nu }^{(-)}$ and $M_{\nu }^{(+)}$ are
the mass matrices for sterile neutrinos. In the limit $\mu \rightarrow 0$,
which corresponds to unbroken lepton number, the active light neutrinos
become massless. The smallness of the $\mu $ parameter yields a small mass
splitting for the two pairs of sterile neutrinos, thus implying that the
sterile neutrinos form pseudo-Dirac pairs.

\begin{figure}[tbp]
\centering
\includegraphics[width=0.5\textwidth]{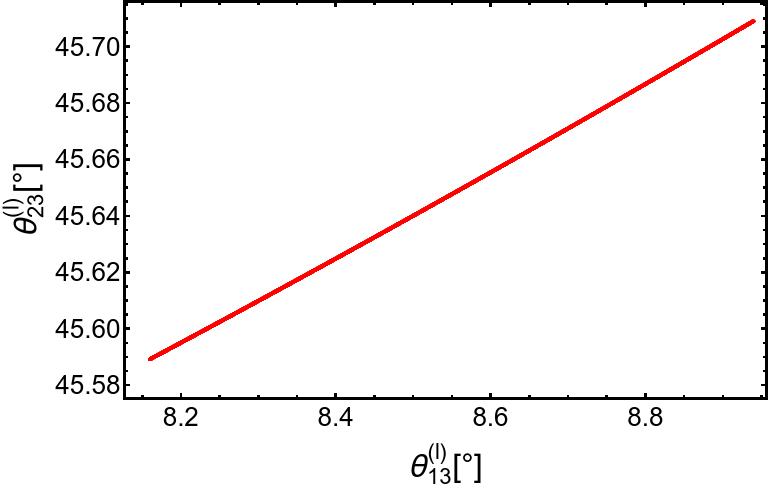}%
\includegraphics[width=0.5\textwidth]{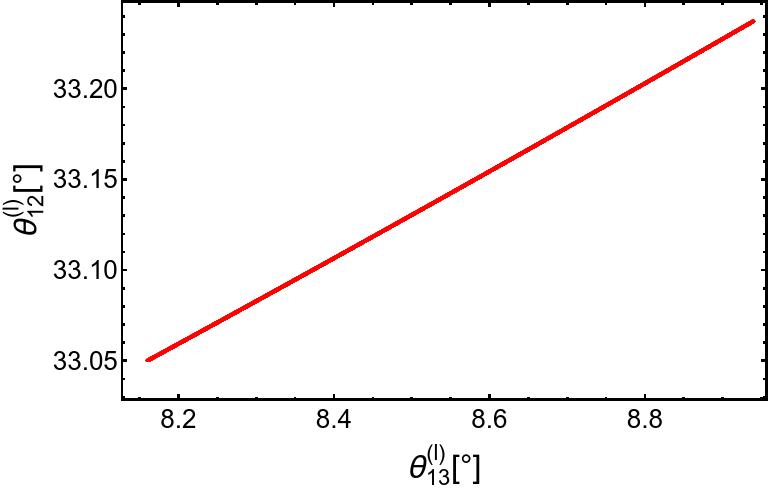}\newline
\includegraphics[width=0.5\textwidth]{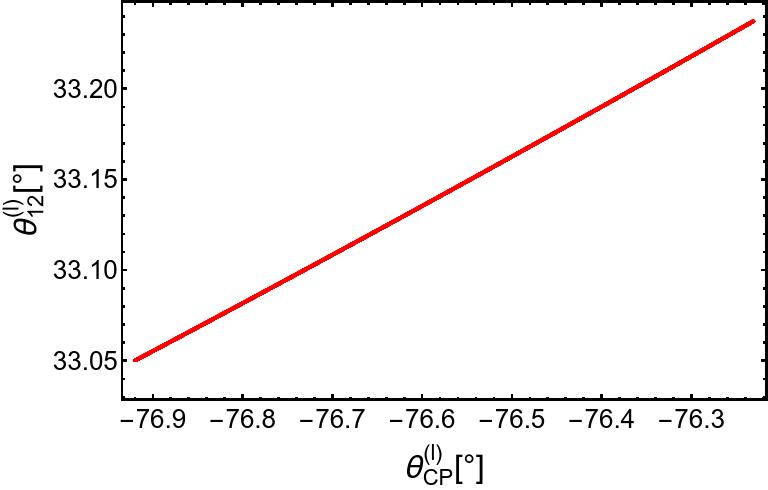}%
\includegraphics[width=0.5\textwidth]{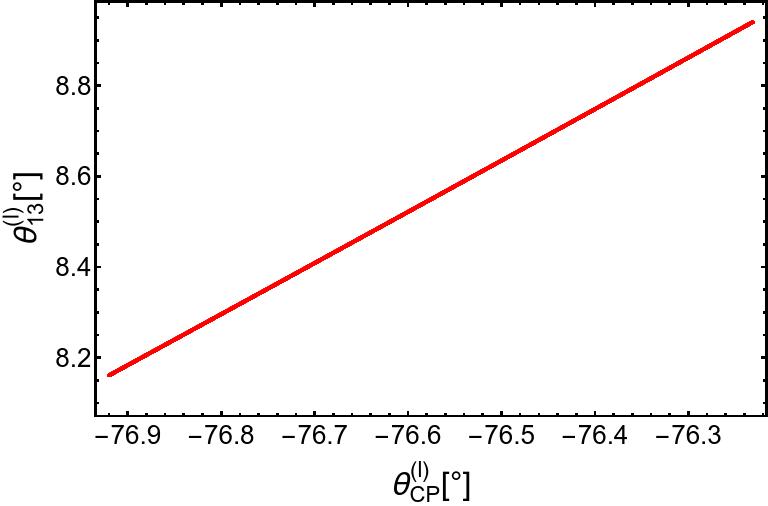}
\caption{Correlations between the different lepton sector observables.}
\label{leptoncorrelations}
\end{figure}

Then, we obtain that the mass matrix for the light active neutrinos has the
form: 
\begin{equation}
\widetilde{M}_{\nu }=\frac{y_{\nu }^{2}v_{u}^{3}v_{d}v_{\varphi _{N}}}{2%
\sqrt{6}y_{\nu N}^{2}v_{\rho }^{2}\Lambda ^{2}}\left( 
\begin{array}{ccc}
\frac{2}{3}\left( p+q\right)  & \frac{p}{\sqrt{2}} & \frac{p}{\sqrt{2}} \\ 
\frac{p}{\sqrt{2}} & q & -\frac{1}{3}\left( p+q\right)  \\ 
\frac{p}{\sqrt{2}} & -\frac{1}{3}\left( p+q\right)  & q%
\end{array}%
\right) =\left( 
\begin{array}{ccc}
\frac{2}{3}\left( A+B\right)  & \frac{A}{\sqrt{2}} & \frac{A}{\sqrt{2}} \\ 
\frac{A}{\sqrt{2}} & B & -\frac{1}{3}\left( A+B\right)  \\ 
\frac{A}{\sqrt{2}} & -\frac{1}{3}\left( A+B\right)  & B%
\end{array}%
\right) ,
\end{equation}%
The light active neutrino mass matrix given above can be diagonalized by a
rotation matrix: 
\begin{equation}
R_{\nu }=\left( 
\begin{array}{ccc}
\frac{\sqrt{\text{$\phi _{g}$}}}{\sqrt[4]{5}} & -\frac{\sqrt{\frac{1}{\text{$%
\phi _{g}$}}}}{\sqrt[4]{5}} & 0 \\ 
\frac{\sqrt{\frac{1}{\text{$\phi _{g}$}}}}{\sqrt{2}\sqrt[4]{5}} & \frac{%
\sqrt{\text{$\phi _{g}$}}}{\sqrt{2}\sqrt[4]{5}} & -\frac{1}{\sqrt{2}} \\ 
\frac{\sqrt{\frac{1}{\text{$\phi _{g}$}}}}{\sqrt{2}\sqrt[4]{5}} & \frac{%
\sqrt{\text{$\phi _{g}$}}}{\sqrt{2}\sqrt[4]{5}} & \frac{1}{\sqrt{2}}%
\end{array}%
\right) =\left( 
\begin{array}{ccc}
\frac{\sqrt{1+\sqrt{5}}}{\sqrt{2}\sqrt[4]{5}} & -\frac{\sqrt{\frac{2}{1+%
\sqrt{5}}}}{\sqrt[4]{5}} & 0 \\ 
\frac{1}{\sqrt[4]{5}\sqrt{1+\sqrt{5}}} & \frac{\sqrt{1+\sqrt{5}}}{2\sqrt[4]{5%
}} & -\frac{1}{\sqrt{2}} \\ 
\frac{1}{\sqrt[4]{5}\sqrt{1+\sqrt{5}}} & \frac{\sqrt{1+\sqrt{5}}}{2\sqrt[4]{5%
}} & \frac{1}{\sqrt{2}} \\ 
&  & 
\end{array}%
\right) ,\hspace{1cm}\hspace{1cm}\text{$\phi _{g}=\frac{1+\sqrt{5}}{2},$}
\end{equation}%
according to the following relation: 
\begin{equation}
R_{\nu }^{T}M_{\nu }R_{\nu }=\left( 
\begin{array}{ccc}
-\frac{\left( 3\sqrt{5}+1\right) A+4B}{\sqrt{6}} & 0 & 0 \\ 
0 & \frac{\left( 3\sqrt{5}-1\right) A-4B}{\sqrt{6}} & 0 \\ 
0 & 0 & -\sqrt{\frac{2}{3}}(A+4B)%
\end{array}%
\right) .
\end{equation}%
Then, the PMNS leptonic mixing matrix has the form: 
\begin{eqnarray}
U &=&R_{l}^{\dagger }R_{\nu }\simeq \left( 
\begin{array}{ccc}
\cos \theta  & 0 & -\sin \theta e^{i\theta } \\ 
0 & 1 & 0 \\ 
\sin \theta e^{-i\theta } & 0 & \cos \theta 
\end{array}%
\right) \left( 
\begin{array}{ccc}
\frac{\sqrt{\text{$\phi _{g}$}}}{\sqrt[4]{5}} & -\frac{\sqrt{\frac{1}{\text{$%
\phi _{g}$}}}}{\sqrt[4]{5}} & 0 \\ 
\frac{\sqrt{\frac{1}{\text{$\phi _{g}$}}}}{\sqrt{2}\sqrt[4]{5}} & \frac{%
\sqrt{\text{$\phi _{g}$}}}{\sqrt{2}\sqrt[4]{5}} & -\frac{1}{\sqrt{2}} \\ 
\frac{\sqrt{\frac{1}{\text{$\phi _{g}$}}}}{\sqrt{2}\sqrt[4]{5}} & \frac{%
\sqrt{\text{$\phi _{g}$}}}{\sqrt{2}\sqrt[4]{5}} & \frac{1}{\sqrt{2}}%
\end{array}%
\right)  \\
&=&\left( 
\begin{array}{ccc}
\frac{1}{5}5^{\frac{3}{4}}\sqrt{\phi _{g}}\cos \theta -\frac{5^{\frac{3}{4}}%
}{10}\sqrt{2}e^{i\theta }\left( \sin \theta \right) \sqrt{\frac{1}{\phi _{g}}%
} & -\frac{1}{5}5^{\frac{3}{4}}\left( \cos \theta \right) \sqrt{\frac{1}{%
\phi _{g}}}-\frac{5^{\frac{3}{4}}}{10}\sqrt{2}\sqrt{\phi _{g}}e^{i\theta
}\sin \theta  & -\frac{1}{2}\sqrt{2}e^{i\theta }\sin \theta  \\ 
\frac{5^{\frac{3}{4}}}{10}\sqrt{2}\sqrt{\frac{1}{\phi _{g}}} & \frac{5^{%
\frac{3}{4}}}{10}\sqrt{2}\sqrt{\phi _{g}} & -\frac{1}{2}\sqrt{2} \\ 
\frac{5^{\frac{3}{4}}}{10}\sqrt{2}\left( \cos \theta \right) \sqrt{\frac{1}{%
\phi _{g}}}+\frac{1}{5}5^{\frac{3}{4}}\sqrt{\phi _{g}}e^{-i\theta }\sin
\theta  & \frac{5^{\frac{3}{4}}}{10}\sqrt{2}\sqrt{\phi _{g}}\cos \theta -%
\frac{1}{5}5^{\frac{3}{4}}e^{-i\theta }\left( \sin \theta \right) \sqrt{%
\frac{1}{\phi _{g}}} & \frac{1}{2}\sqrt{2}\cos \theta 
\end{array}%
\right) \allowbreak   \notag
\end{eqnarray}%
The experimental values of the SM charged lepton masses: 
\begin{equation*}
m_{e}(m_{Z})=0.4883266\pm 0.0000017MeV,\ \ m_{\mu }(m_{Z})=102.87267\pm
0.00021MeV,\ \ m_{\tau }(m_{Z})=1747.43\pm 0.12MeV,
\end{equation*}%
neutrino mass squared splittings, leptonic mixing angles and leptonic Dirac
CP violating phase, as shown in Table \ref{tab:neutrinos-NH}, can be very
well reproduced for the following bechmark point: 
\begin{equation}
M_{l}=\left( 
\begin{array}{ccc}
0.000485825 & 0 & 0.0343002\,-0.366171i \\ 
0.0000470777 & 0.101891 & 0 \\ 
0 & 0 & 1.70203 \\ 
&  & 
\end{array}%
\right) \mathit{GeV},\hspace{1cm}A=0.236\mathit{meV},\hspace{1cm}B=17.82%
\mathit{meV}
\end{equation}%
\begin{table}[t]
{\footnotesize \ }
\par
\begin{center}
{\footnotesize \ \renewcommand{\arraystretch}{1} 
\begin{tabular}{c|c||c|c|c|c}
\hline
\multirow{2}{*}{Observable} & \multirow{2}{*}{\parbox{7em}{Model\\ value} }
& \multicolumn{4}{|c}{Neutrino oscillation global fit values (NH)} \\ 
\cline{3-6}
&  & Best fit $\pm 1\sigma$ \cite{deSalas:2020pgw} & Best fit $\pm 1\sigma$ 
\cite{Esteban:2020cvm} & $3\sigma$ range \cite{deSalas:2020pgw} & $3\sigma$
range \cite{Esteban:2020cvm} \\ \hline\hline
$\Delta m_{21}^{2}$ [$10^{-5}$eV$^{2}$] & $7.50$ & $7.50_{-0.20}^{+0.22}$ & $%
7.42_{-0.20}^{+0.21}$ & $6.94-8.14$ & $6.82-8.04$ \\ \hline
$\Delta m_{31}^{2}$ [$10^{-3}$eV$^{2}$] & $2.56$ & $2.56_{-0.04}^{+0.03}$ & $%
2.517_{-0.028}^{+0.026}$ & $2.46-2.65$ & $2.435-2.598$ \\ \hline
$\theta _{12}(^{\circ })$ & $33.15$ & $34.3\pm 1.0$ & $33.44_{-0.74}^{+0.77}$
& $31.4-37.4$ & $31.27-35.86$ \\ \hline
$\theta _{13}(^{\circ })$ & $8.59$ & $8.58_{-0.15}^{+0.11}$ & $8.57\pm 0.12$
& $8.16-8.94$ & $8.20-8.93$ \\ \hline
$\theta _{23}(^{\circ })$ & $45.65$ & $48.79_{-1.25}^{+0.93}$ & $%
49.2_{-1.2}^{+0.9} $ & $41.63-51.32$ & $40.1-51.7$ \\ \hline
$\delta _{CP}(^{\circ })$ & $-76.54$ & $216_{-25}^{+41}$ & $197_{-24}^{+27}$
& $144-360$ & $120-369$ \\ \hline\hline
\end{tabular}
{\normalsize \ }}
\end{center}
\par
{\footnotesize \ }
\caption{Model and experimental values of the neutrino mass squared
splittings, leptonic mixing angles, and $CP$ -violating phase. The
experimental values are taken from Refs.~\protect\cite%
{deSalas:2020pgw,Esteban:2020cvm}.}
\label{tab:neutrinos-NH}
\end{table}
As shown in Figure \ref{leptoncorrelations}, there is a linear correlation
between the different lepton sector observables. As shown in Table \ref%
{tab:neutrinos-NH} and Figure \ref{leptoncorrelations}, our model is
consistent with the neutrino oscillation experimental data. 

\newpage

\section{Conclusions \label{sec:conc}}

We have built a supersymmetric model where leptonic mixing consists in a
perturbed golden ratio mixing pattern. The model is based on an $A_{5}$
family symmetry, supplemented by the $Z_{9}\times Z_{N}$ discrete group. The
tiny masses of the light active neutrinos are generated by an inverse seesaw
mechanism mediated by right handed Majorana neutrinos. In this framework,
renormalizable and non-renormalizable Majorana mass terms generated after
the spontaneous breaking of the discrete symmetries. The inverse seesaw
mechanism is guaranteed through the smallness of the $\mu $ parameter, which
arises from non-renormalizable Yukawa terms involving gauge singlet Majorana
neutrinos. The resulting mass matrix for the light active neutrinos features
a golden ratio mixing pattern in the family symmetry basis, in this same
basis the charged lepton mass matrix is not diagonal. Therefore, the charged
lepton sector provides the required perturbation, leading to a leptonic
mixing pattern in excellent agreement with the experimental data.
Furthermore, our model predicts linear correlations among the leptonic
mixing angles and the Dirac leptonic CP violating phase.

\section*{Acknowledgments}

A.E.C.H is supported by ANID-Chile FONDECYT 1210378, ANID PIA/APOYO
AFB180002 and ANID- Programa Milenio - code ICN2019\_044. IdMV acknowledges
funding from Funda\c{c}\~{a}o para a Ci\^{e}ncia e a Tecnologia (FCT)
through the contract UID/FIS/00777/2020 and was supported in part by FCT
through projects CFTP-FCT Unit 777 (UID/FIS/00777/2019),
PTDC/FIS-PAR/29436/2017, CERN/FIS-PAR/0004/2019 and CERN/FIS-PAR/0008/2019
which are partially funded through POCTI (FEDER), COMPETE, QREN and EU. AECH
thanks the Instituto Superior T\'{e}cnico, Universidade de Lisboa for
hospitality, where part of this work was done.

\appendix

\section*{Appendix A. $A_5$ tensor product rules}

In this appendix we show the tensor product rules of the $A_5$ discrete
group. We assign $a=(a_1,a_2,a_3)^T$ and $b=(b_1,b_2,b_3)^T$ to the \textbf{3%
} representation, while $a^{\prime }=(a_1^{\prime },a_2^{\prime
},a_3^{\prime T}$ and $b^{\prime }=(b_1^{\prime },b_2^{\prime },b_3^{\prime
T}$ belong to the $\mathbf{3^{\prime }}$ representation. $%
c=(c_1,c_2,c_3,c_4,c_5)^T$ and $d=(d_1,d_2,d_3,d_4,d_5)^T$ are pentaplets; $%
f=(f_1,f_2,f_3,f_4)^T$ and $g=(g_1,g_2,g_3,g_4)^T$ are tetraplets. \vspace{%
0.3cm} \newline
$\mathbf{3} \,\otimes\, \mathbf{3}=~\mathbf{3}_a~+~(\mathbf{1}~+~\mathbf{5}%
)_s\hfill$ 
\begin{eqnarray}
\mathbf{1}&=& a_1 b_1+a_2 b_3+a_3 b_2  \notag \\
\mathbf{3}&=& (a_2 b_3 - a_3 b_2,~a_1 b_2 - a_2 b_1,~a_3 b_1-a_1 b_3)^T 
\notag \\
\mathbf{5}&=& (a_1 b_1-\frac{a_2 b_3}{2}-\frac{a_3 b_2}{2},\frac{\sqrt{3}}{2}%
(a_1 b_2 + a_2 b_1), -\sqrt{\frac{3}{2}} a_2 b_2, -\sqrt{\frac{3}{2}} a_3
b_3, -\frac{\sqrt{3}}{2}(a_1 b_3 + a_3 b_1))^T  \notag
\end{eqnarray}
\newline
$\mathbf{3^{\prime }}\,\otimes\, \mathbf{3^{\prime }}=~\mathbf{3^{\prime }}%
_a~+~(\mathbf{1}~+~\mathbf{5})_s\hfill$ 
\begin{eqnarray}
\mathbf{1}&=& a^{\prime }_1 b^{\prime }_1+a^{\prime }_2 b^{\prime
}_3+a^{\prime }_3 b^{\prime }_2  \notag \\
\mathbf{3^{\prime }}&=& (a^{\prime }_2 b^{\prime }_3 - a^{\prime }_3
b^{\prime }_2,~a^{\prime }_1 b^{\prime }_2 - a^{\prime }_2 b^{\prime
}_1,~a^{\prime }_3 b^{\prime }_1-a^{\prime }_1 b^{\prime }_3)^T  \notag \\
\mathbf{5}&=& (a^{\prime }_1 b^{\prime }_1-\frac{a^{\prime }_2 b^{\prime }_3%
}{2}-\frac{a^{\prime }_3 b^{\prime }_2}{2},\sqrt{\frac{3}{2}} a^{\prime }_3
b^{\prime }_3,-\frac{\sqrt{3}}{2}(a^{\prime }_1 b^{\prime }_2 + a^{\prime
}_2 b^{\prime }_1),-\frac{\sqrt{3}}{2}(a^{\prime }_1 b^{\prime }_3 +
a^{\prime }_3 b^{\prime }_1), -\sqrt{\frac{3}{2}} a^{\prime }_2 b^{\prime
}_2)^T  \notag
\end{eqnarray}
\newline
$\mathbf{3}\,\otimes\, \mathbf{3^{\prime }} =~\mathbf{4}~+~\mathbf{5}\hfill$ 
\begin{eqnarray}
\mathbf{4}&=& (a_2 b^{\prime }_1 - \frac{a_3 b^{\prime }_2}{\sqrt{2}},-a_1
b^{\prime }_2 + \frac{a_3 b^{\prime }_3}{\sqrt{2}},a_1 b^{\prime }_3 - \frac{%
a_2 b^{\prime }_2}{\sqrt{2}}, -a_3 b^{\prime }_1 + \frac{a_2 b^{\prime }_3}{%
\sqrt{2}})^T  \notag \\
\mathbf{5}&=& (a_1 b^{\prime }_1,-\frac{a_2 b^{\prime }_1 + \sqrt{2}a_3
b^{\prime }_2}{\sqrt{3}},\frac{a_1 b^{\prime }_2 + \sqrt{2}a_3 b^{\prime }_3%
}{\sqrt{3}}, \frac{a_1 b^{\prime }_3 + \sqrt{2}a_2 b^{\prime }_2}{\sqrt{3}},%
\frac{a_3 b^{\prime }_1 + \sqrt{2}a_2 b^{\prime }_3}{\sqrt{3}}))^T  \notag
\end{eqnarray}
\newline
$\mathbf{3}\,\otimes\, \mathbf{4}\,~=~\mathbf{3^{\prime }}~+~\mathbf{4}~+~%
\mathbf{5}\hfill$ 
\begin{eqnarray}
\mathbf{3^{\prime }}&=& (a_2 g_4 -a_3 g_1,\frac{1}{\sqrt{2}}(\sqrt{2} a_1
g_2 +a_2 g_1 +a_3 g_3),-\frac{1}{\sqrt{2}}(\sqrt{2} a_1 g_3 +a_2 g_2 +a_3
g_4))^T  \notag \\
\mathbf{4}&=& (a_1 g_1 + \sqrt{2} a_3 g_2,-a_1 g_2 + \sqrt{2} a_2 g_1,a_1
g_3 - \sqrt{2} a_3 g_4, -a_1 g_4 - \sqrt{2} a_2 g_3)^T  \notag \\
\mathbf{5}&=& (a_3 g_1 +a_2 g_4,\sqrt{\frac{2}{3}}(\sqrt{2} a_1 g_1-a_3 g_2),%
\frac{1}{\sqrt{6}}(\sqrt{2}a_1 g_2-3 a_3 g_3+a_2 g_1 ),  \notag \\
& & \frac{1}{\sqrt{6}}(\sqrt{2}a_1 g_3-3 a_2 g_2+a_3 g_4 ),\sqrt{\frac{2}{3}}%
(-\sqrt{2} a_1 g_4+a_2 g_3))^T  \notag
\end{eqnarray}
\newline
$\mathbf{3^{\prime }}\,\otimes\, \mathbf{4}\,~=~\mathbf{3}~+~\mathbf{4}~+~%
\mathbf{5}\hfill$ 
\begin{eqnarray}
\mathbf{3}&=& (a^{\prime }_2 g_3 -a^{\prime }_3 g_2,\frac{1}{\sqrt{2}}(\sqrt{%
2} a^{\prime }_1 g_1 +a^{\prime }_2 g_4-a^{\prime }_3 g_3),\frac{1}{\sqrt{2}}%
(-\sqrt{2} a^{\prime }_1 g_4 +a^{\prime }_2 g_2 -a^{\prime }_3 g_1))^T 
\notag \\
\mathbf{4}&=& (a^{\prime }_1 g_1 + \sqrt{2} a^{\prime }_3 g_3,a^{\prime }_1
g_2 - \sqrt{2} a^{\prime }_3 g_4,-a^{\prime }_1 g_3 + \sqrt{2} a^{\prime }_2
g_1, -a^{\prime }_1 g_4 - \sqrt{2} a^{\prime }_2 g_2)^T  \notag \\
\mathbf{5}&=& (a^{\prime }_3 g_2 +a^{\prime }_2 g_3,\frac{1}{\sqrt{6}}(\sqrt{%
2}a^{\prime }_1 g_1 -3 a^{\prime }_2 g_4-a^{\prime }_3 g_3 ),-\sqrt{\frac{2}{%
3}}(\sqrt{2} a^{\prime }_1 g_2+a^{\prime }_3 g_4),  \notag \\
& & -\sqrt{\frac{2}{3}}(\sqrt{2} a^{\prime }_1 g_3+a^{\prime }_2 g_1),\frac{1%
}{\sqrt{6}}(-\sqrt{2}a^{\prime }_1 g_4+3 a^{\prime }_3 g_1+a^{\prime }_2 g_2
))^T  \notag
\end{eqnarray}
\newline
\newline
$\mathbf{3 }\,\otimes\, \mathbf{5}\,~=~\mathbf{3}~+~\mathbf{3^{\prime }}~+~%
\mathbf{4}~+~\mathbf{5}\hfill$ 
\begin{eqnarray}
\mathbf{3}&=&(\frac{2 a_1 c_1}{\sqrt{3}}+ a_3 c_2 -a_2 c_5,-\frac{ a_2 c_1}{%
\sqrt{3}}+ a_1 c_2 -\sqrt{2}a_3 c_3, -\frac{ a_3 c_1}{\sqrt{3}}- a_1 c_5 -%
\sqrt{2}a_2 c_4)^T  \notag \\
\mathbf{3^{\prime }}&=&( a_1 c_1+\frac{a_2 c_5-a_3 c_2}{\sqrt{3}},\frac{ a_1
c_3 +\sqrt{2}(a_3 c_4-a_2 c_2)}{\sqrt{3}}, \frac{ a_1 c_4 +\sqrt{2}(a_2 c_3
+ a_3 c_5)}{\sqrt{3}})^T  \notag \\
\mathbf{4}&=&(4 a_1 c_2 +2\sqrt{3}a_2 c_1 +\sqrt{2}a_3 c_3,2 a_1 c_3 -2\sqrt{%
2}a_2 c_2 -3\sqrt{2}a_3 c_4,  \notag \\
& & 2 a_1 c_4 -3\sqrt{2}a_2 c_3 +2 \sqrt{2}a_3 c_5,-4 a_1 c_5 +\sqrt{2}a_2
c_4 +2\sqrt{3}a_3 c_1)^T  \notag \\
\mathbf{5}&=&(a_2 c_5 +a_3 c_2,a_2 c_1 -\frac{a_1 c_2 +\sqrt{2}a_3 c_3}{%
\sqrt{3}},-\frac{2 a_1 c_3 +\sqrt{2}a_2 c_2}{\sqrt{3}},  \notag \\
& & \frac{2 a_1 c_4 -\sqrt{2}a_3 c_5}{\sqrt{3}},a_3 c_1 +\frac{a_1 c_5 -%
\sqrt{2}a_2 c_4}{\sqrt{3}})^T  \notag
\end{eqnarray}
\newline
$\mathbf{3^{\prime }}\,\otimes\, \mathbf{5}\,~=~\mathbf{3}~+~\mathbf{%
3^{\prime }}~+~\mathbf{4}~+~\mathbf{5}\hfill$ 
\begin{eqnarray}
\mathbf{3}&=&( a^{\prime }_1 c_1+\frac{a^{\prime }_3 c_3+a^{\prime }_2 c_4}{%
\sqrt{3}},\frac{- a^{\prime }_1 c_2 +\sqrt{2}(a^{\prime }_3 c_4+a^{\prime
}_2 c_5)}{\sqrt{3}}, \frac{ a^{\prime }_1 c_5 +\sqrt{2}(a^{\prime }_2 c_3 -
a^{\prime }_3 c_2)}{\sqrt{3}})^T  \notag \\
\mathbf{3^{\prime }}&=&(\frac{2 a^{\prime }_1 c_1}{\sqrt{3}}- a^{\prime }_3
c_3 -a^{\prime }_2 c_4,-\frac{ a^{\prime }_2 c_1}{\sqrt{3}}- a^{\prime }_1
c_3 -\sqrt{2}a^{\prime }_3 c_5, -\frac{ a^{\prime }_3 c_1}{\sqrt{3}}-
a^{\prime }_1 c_4 +\sqrt{2}a^{\prime }_2 c_2)^T  \notag \\
\mathbf{4}&=&(2 a^{\prime }_1 c_2 + 3\sqrt{2}a^{\prime }_2 c_5 -2\sqrt{2}%
a^{\prime }_3 c_4,-4 a^{\prime }_1 c_3 +2\sqrt{3}a^{\prime }_2 c_1 +\sqrt{2}%
a^{\prime }_3 c_5,  \notag \\
& & -4 a^{\prime }_1 c_4 -\sqrt{2}a^{\prime }_2 c_2 +2\sqrt{3}a^{\prime }_3
c_1 ,-2 a^{\prime }_1 c_5 -2\sqrt{2}a^{\prime }_2 c_3 - 3\sqrt{2}a^{\prime
}_3 c_2)^T  \notag \\
\mathbf{5}&=&(a^{\prime }_2 c_4 -a^{\prime }_3 c_3,\frac{2 a^{\prime }_1 c_2
+\sqrt{2}a^{\prime }_3 c_4}{\sqrt{3}},-a^{\prime }_2 c_1 -\frac{a^{\prime
}_1 c_3 -\sqrt{2}a^{\prime }_3 c_5}{\sqrt{3}},  \notag \\
& & a^{\prime }_3 c_1 +\frac{a^{\prime }_1 c_4 +\sqrt{2}a^{\prime }_2 c_2}{%
\sqrt{3}},\frac{-2 a^{\prime }_1 c_5 +\sqrt{2}a^{\prime }_2 c_3}{\sqrt{3}})^T
\notag
\end{eqnarray}
\newline
$~\,\mathbf{4}\,\otimes\, \mathbf{4}\,~=~(\mathbf{3}~ +~\mathbf{3^{\prime }}%
)_a~+~(\mathbf{1}~+~\mathbf{4}~ +~\mathbf{5})_s\hfill$ 
\begin{eqnarray}
\mathbf{1}&=& f_1 g_4 +f_2 g_3 +f_3 g_2 +f_4 g_1  \notag \\
\mathbf{3}&=&(f_1 g_4 -f_4 g_1 +f_3 g_2- f_2 g_3,\sqrt{2}(f_2 g_4 -f_4 g_2), 
\sqrt{2}(f_1 g_3 -f_3 g_1))^T  \notag \\
\mathbf{3^{\prime }}&=&(f_1 g_4 -f_4 g_1 +f_2 g_3- f_3 g_2,\sqrt{2}(f_3 g_4
-f_4 g_3), \sqrt{2}(f_1 g_2 -f_2 g_1))^T  \notag \\
\mathbf{4}&=&(f_3 g_3 -f_4 g_2 -f_2 g_4,f_1 g_1 +f_3 g_4 +f_4 g_3, -f_4 g_4
-f_1 g_2 -f_2 g_1,-f_2 g_2 +f_1 g_3 +f_3 g_1)^T  \notag \\
\mathbf{5}&=&(f_1 g_4 +f_4 g_1 -f_3 g_2 -f_2 g_3,-\sqrt{\frac{2}{3}}(2 f_3
g_3 +f_2 g_4 +f_4 g_2), \sqrt{\frac{2}{3}}(-2 f_1 g_1 +f_3 g_4 +f_4 g_3), 
\notag \\
& & \sqrt{\frac{2}{3}}(-2 f_4 g_4 +f_2 g_1 +f_1 g_2),\sqrt{\frac{2}{3}}(2
f_2 g_2 +f_1 g_3 +f_3 g_1) )^T  \notag
\end{eqnarray}
\newline
$~\,\mathbf{4}\,\otimes\, \mathbf{5}\,~=~\mathbf{3}~ +~\mathbf{3^{\prime }}%
~+~\mathbf{4}~+~\mathbf{5}~ +~\mathbf{5}\hfill$ 
\begin{eqnarray}
\mathbf{3}&=&(4 f_1 c_5 -4 f_4 c_2 -2 f_3 c_3-2 f_2 c_4,-2 \sqrt{3} f_1 c_1-%
\sqrt{2}(2 f_2 c_5 -3 f_3 c_4+f_4 c_3),  \notag \\
& & \sqrt{2} (-f_1 c_4+ 3 f_2 c_3 +2 f_3 c_2)-2\sqrt{3} f_4 c_1)^T  \notag \\
\mathbf{3^{\prime }}&=&(2 f_1 c_5 -2 f_4 c_2 +4 f_3 c_3+4 f_2 c_4,-2 \sqrt{3}
f_2 c_1+\sqrt{2}(2 f_4 c_4 +3 f_1 c_2-f_3 c_5),  \notag \\
& & \sqrt{2} (f_2 c_2- 3 f_4 c_5 +2 f_1 c_3)-2\sqrt{3} f_3 c_1)^T  \notag \\
\mathbf{4}&=&(3 f_1 c_1+\sqrt{6}(f_2 c_5 +f_3 c_4-2 f_4 c_3),-3 f_2 c_1+%
\sqrt{6}(f_4 c_4 -f_1 c_2 +2 f_3 c_5),  \notag \\
& & -3 f_3 c_1+\sqrt{6}(f_1 c_3 +f_4 c_5-2 f_2 c_2),3 f_4 c_1+\sqrt{6}(f_2
c_3 -f_3 c_2-2 f_1 c_4))^T  \notag \\
\mathbf{5_1}&=&(f_1 c_5 +2 f_2 c_4 -2 f_3 c_3+f_4 c_2,-2 f_1 c_1+\sqrt{6}
f_2 c_5 ,f_2 c_1+\sqrt{\frac{3}{2}}(-f_1 c_2 -f_3 c_5+2 f_4 c_4),  \notag \\
& & -f_3 c_1-\sqrt{\frac{3}{2}}(f_2 c_2 +f_4 c_5+2 f_1 c_3),-2 f_4 c_1-\sqrt{%
6} f_3 c_2)^T  \notag \\
\mathbf{5_2}&=&(f_2 c_4 - f_3 c_3,-f_1 c_1+\frac{2 f_2 c_5-f_3 c_4 -f_4 c_3}{%
\sqrt{6}},-\sqrt{\frac{2}{3}}(f_1 c_2 +f_3 c_5- f_4 c_4),  \notag \\
& & -\sqrt{\frac{2}{3}}(f_1 c_3 +f_2 c_2+ f_4 c_5),-f_4 c_1-\frac{2 f_3
c_2+f_1 c_4 +f_2 c_3}{\sqrt{6}})^T  \notag
\end{eqnarray}
\newline
$~\,\mathbf{5}\,\otimes\, \mathbf{5}\,~=~(\mathbf{3}~ +~\mathbf{3^{\prime }}%
~ +~\mathbf{4})_a~ +~(\mathbf{1}~+~\mathbf{4}~ +~\mathbf{5}~ +~\mathbf{5}%
)_s\hfill$ 
\begin{eqnarray}
\mathbf{3}&=&(2 ( c_4 d_3 -c_3 d_4)+c_2 d_5 -c_5 d_2,\sqrt{3}(c_2 d_1-c_1
d_2)+\sqrt{2}(c_3 d_5 -c_5 d_3),  \notag \\
& & \sqrt{3}(c_5 d_1-c_1 d_5)+\sqrt{2}(c_4 d_2 -c_2 d_4))^T  \notag \\
\mathbf{3^{\prime }}&=&(2 ( c_2 d_5 -c_5 d_2)+c_3 d_4 -c_4 d_3,\sqrt{3}(c_3
d_1-c_1 d_3)+\sqrt{2}(c_4 d_5 -c_5 d_4),  \notag \\
& & \sqrt{3}(c_1 d_4-c_4 d_1)+\sqrt{2}(c_3 d_2 -c_2 d_3))^T  \notag \\
\mathbf{4_s}&=& ((c_1 d_2 +c_2 d_1)-\frac{(c_3 d_5 +c_5 d_3) -4 c_4 d_4}{%
\sqrt{6}},-(c_1 d_3 + c_3 d_1)-\frac{(c_4 d_5 +c_5 d_4) -4 c_2 d_2}{\sqrt{6}}%
,  \notag \\
& & (c_1 d_4 +c_4 d_1)-\frac{(c_2 d_3 +c_3 d_2) +4 c_5 d_5}{\sqrt{6}},(c_1
d_5 + c_5 d_1)-\frac{(c_2 d_4 +c_4 d_2) +4 c_3 d_3}{\sqrt{6}})^T  \notag \\
\mathbf{4_a}&=& ((c_1 d_2 -c_2 d_1)+\sqrt{\frac{3}{2}}(c_3 d_5-c_5 d_3),(c_1
d_3 - c_3 d_1)+\sqrt{\frac{3}{2}}(c_4 d_5-c_5 d_4),  \notag \\
& & (c_4 d_1 -c_1 d_4)+\sqrt{\frac{3}{2}}(c_3 d_2-c_2 d_3),(c_1 d_5 - c_5
d_1)+\sqrt{\frac{3}{2}}(c_4 d_2-c_2 d_4))^T  \notag \\
\mathbf{5_1}&=& (c_1 d_1 +c_2 d_5 +c_5 d_2 +\frac{c_3 d_4 +c_4 d_3}{2},
-(c_1 d_2 +c_2 d_1)+ \sqrt{\frac{3}{2}} c_4 d_4, \frac{1}{2}(c_1 d_3 + c_3
d_1-\sqrt{6}(c_4 d_5 +c_5 d_4)),  \notag \\
& & \frac{1}{2}(c_1 d_4 + c_4 d_1+\sqrt{6}(c_2 d_3 +c_3 d_2)),-(c_1 d_5 +c_5
d_1) -\sqrt{\frac{3}{2}} c_3 d_3)^T  \notag \\
\mathbf{5_2}&=& (\frac{2 c_1 d_1 +c_2 d_5 +c_5 d_2}{2}, \frac{-3(c_1 d_2+c_2
d_1)+\sqrt{6}(2 c_4 d_4 +c_3 d_5+c_5 d_3)}{6}, -\frac{2 c_4 d_5 +2 c_5
d_4+c_2 d_2}{\sqrt{6}},  \notag \\
& & \frac{2 c_2 d_3 +2 c_3 d_2-c_5 d_5}{\sqrt{6}},\frac{-3(c_1 d_5+c_5 d_1)+%
\sqrt{6}(-2 c_3 d_3 +c_2 d_4+c_4 d_2)}{6})^T  \notag \\
\mathbf{1}&=& c_1 d_1+c_3 d_4+c_4 d_3-c_2 d_5 -c_5 d_2  \notag
\end{eqnarray}

\bibliographystyle{utphys}
\bibliography{A5Refs.bib}

\end{document}